\documentclass[iop]{emulateapj}
\slugcomment{{\sc Accepted to ApJ:} 7/21/2015}
\usepackage{graphicx}
\usepackage{natbib}
\usepackage{latexsym}
\usepackage{amssymb}
\usepackage{longtable}
\usepackage{amsmath}
\usepackage{url}
\def\msun{\ifmmode {\rm\,M_\odot}\else ${\rm\,M_\odot}$\fi}
\def\Msun{\ifmmode {\rm\,\it{M_\odot}}\else ${\rm\,M_\odot}$\fi}
\def\lsun{\ifmmode {\rm\,L_\odot}\else ${\rm\,L_\odot}$\fi}
\def\Lsun{\ifmmode {\rm\,\it{L_\odot}}\else ${\rm\,L_\odot}$\fi}
\def\rsun{\ifmmode {\rm\,R_\odot}\else ${\rm\,R_\odot}$\fi}
\def\Rsun{\ifmmode {\rm\,\it{R_\odot}}\else ${\rm\,R_\odot}$\fi}
\def\Tsun{\ifmmode {\rm\,T_\odot}\else ${\rm\,T_\odot}$\fi}
\def\arcsec{\ifmmode {^{\prime\prime}}\else $^{\prime\prime}$\fi}
\def\asec{\ifmmode {^{\prime\prime}}\else $^{\prime\prime}$\fi}
\def\arcmin{\ifmmode {^{\prime}}\else $^{\prime}$\fi}
\def\amin{\ifmmode {^{\prime}}\else $^{\prime}$\fi}
\def\simlt{\mathrel{\spose{\lower 3pt\hbox{$\mathchar"218$}}
     \raise 2.0pt\hbox{$\mathchar"13C$}}}
\def\simgt{\mathrel{\spose{\lower 3pt\hbox{$\mathchar"218$}}
\     \raise 2.0pt\hbox{$\mathchar"13E$}}}

\usepackage[pdftex,backref,breaklinks,colorlinks,citecolor=blue]{hyperref}
\usepackage[all]{hypcap}

\begin{document}

\title{Optical Hydrogen Absorption Consistent with a Thin Bow Shock leading the Hot Jupiter HD 189733b}

\author{P. Wilson Cauley and Seth Redfield}
\email{pcauley@wesleyan.edu}
\affil{Wesleyan University}
\affil{Astronomy Department, Van Vleck Observatory, 96 Foss Hill Dr., Middletown, CT 06459}

\author{Adam G. Jensen}
\affil{University of Nebraska-Kearney}
\affil{Department of Physics \& Physical Science, 24011 1th Avenue, Kearney, NE 68849}

\author{Travis Barman}
\affil{University of Arizona}
\affil{Department of Planetary Sciences and Lunar and Planetary Laboratory, 1629 E University Blvd., Tuscon, AZ 85721}

\author{Michael Endl and William D. Cochran}
\affil{The University of Texas at Austin}
\affil{Department of Astronomy and McDonald Observatory, 2515 Speedway, C1400, Austin, TX 78712}

\begin{abstract} 

Bow shocks are ubiquitous astrophysical phenomena resulting from the supersonic passage of an object through a gas. Recently, pre-transit absorption in UV metal transitions of the hot Jupiter exoplanets HD 189733b and WASP12-b have been interpreted as being caused by material compressed in a planetary bow shock. Here we present a robust detection of a time-resolved pre-transit, as well as in-transit, absorption signature around the hot Jupiter exoplanet HD 189733b using high spectral resolution observations of several hydrogen Balmer lines. The line shape of the pre-transit feature and the shape of the time series absorption provide the strongest constraints on the morphology and physical characteristics of extended structures around an exoplanet. The in-transit measurements confirm the previous exospheric H$\alpha$ detection although the absorption depth measured here is $\sim$50\% lower. The pre-transit absorption feature occurs 125 minutes before the predicted optical transit, a projected linear distance from the planet to the stellar disk of 7.2 $R_p$. The absorption strength observed in the Balmer lines indicates an optically thick, but physically small, geometry. We model this signal as the early ingress of a planetary bow shock. If the bow shock is mediated by a planetary magnetosphere, the large standoff distance derived from the model suggests a large planetary magnetic field strength of $B_{eq}$=28 G. Better knowledge of exoplanet magnetic field strengths is crucial to understanding the role these fields play in planetary evolution and the potential development of life on planets in the habitable zone. 

\end{abstract}


\section{INTRODUCTION}

Hot Jupiters (HJs) are Jupiter-mass planets that are orbiting within $\sim$10 stellar radii of their host stars and have orbital periods of a few days. The large insolation of these planets results in extreme heating of their bound atmospheres and inflated radii \citep{burrows}. These objects provide a view into planetary atmospheric physics that do not occur in our solar system. Some HJs are believed to be losing mass via hydrodynamic blowoff \citep{vidal}. The close proximity of a HJ to its parent star can result in star-planet-interactions (SPIs) \citep{cuntz,shkolnik} and, potentially, the formation of bow shocks around the planet as it moves supersonically through the stellar wind or corona \citep{vidotto10}. If the bow shock around a HJ is mediated by the planetary magnetosphere, absorption by material compressed by the shock can be used to estimate the strength of the planetary magnetic field. 

One of the closest of these HJs to our solar system, HD 189733b \citep{bouchy}, has been studied extensively due to the relative brightness of its host star. Various molecules and atoms in both the bound and unbound atmosphere of HD 189733b, including (but not limited to) CH$_4$ \citep{swain08}, H$_2$0 \citep{grillmair}, Na I \citep{redfield}, O I \citep{benjaffel}, and neutral hydrogen \citep{jensen12,desetangs,desetangs12}, have been detected using transmission spectroscopy, where stellar photons are absorbed by the planetary atmosphere while the planet transits its host star.

In addition to in-transit absorption, hints of pre-transit absorption have been observed in HD 189733b \citep{benjaffel,bourrier13} and WASP-12b \citep{fossati} and provided early indications that material exterior to an exoplanet's atmosphere could have sufficient optical depth to cause an absorption signature. The WASP-12b result, a single 2--sigma absorption of a 41 \AA \hspace{0pt} region in the near-UV, was interpreted and modeled as a transiting bow shock \citep{vidotto10,llama11}. \citet{bourrier13} measured pre--transit absorption in Si III 1206.5 \AA\hspace{0pt} for one of two analyzed HD 189733b transits. \citet{benjaffel} report two pre-transit UV absorption measurements for HD 189733b, observed in C II, but not O I, and they model the result as a transiting bow shock \citep{benjaffel,llama13}. 

\begin{figure*}[htbp]\label{fig:fig1}
   \centering
   \includegraphics[scale=.65,clip,trim=5mm 25mm 10mm 30mm,angle=0]{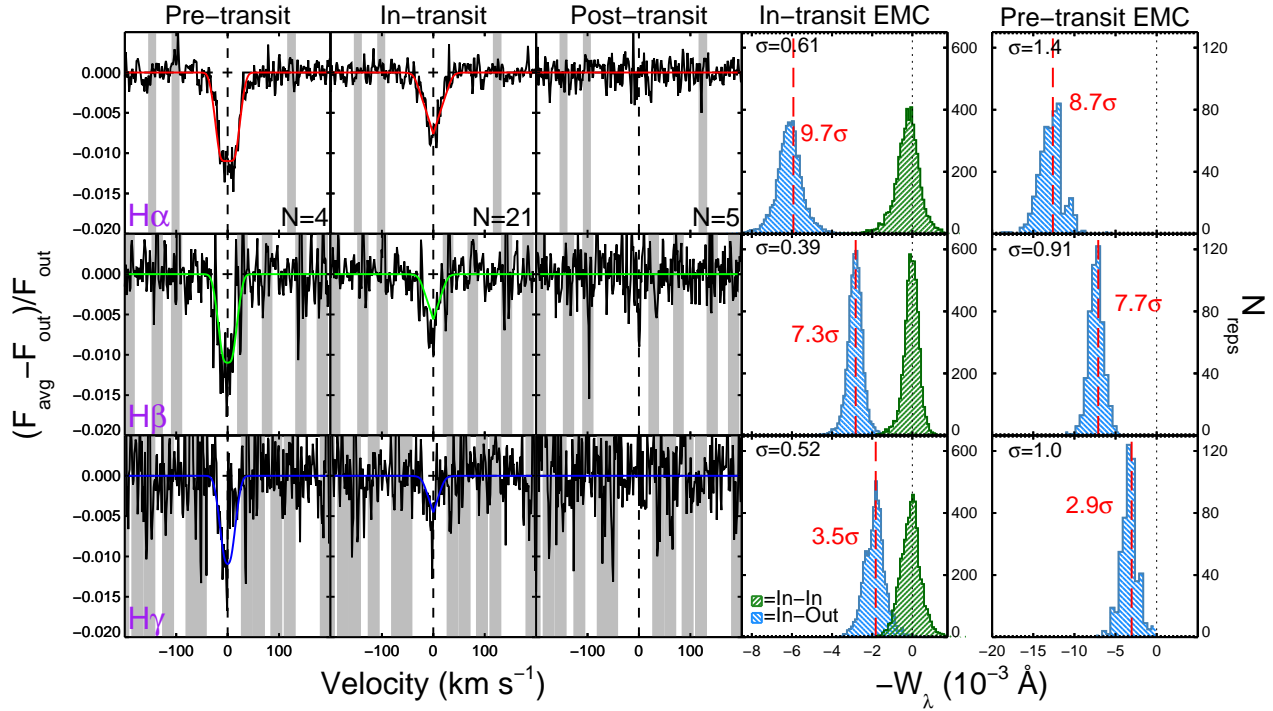} 
   \caption{The averaged transmission spectra for the pre-, in-, and post-transit spectra for H$\alpha$ (top row), H$\beta$ (middle row), and H$\gamma$ (bottom row). The cores of identified stellar lines other than the lines of interest are masked in gray. These points are not included in the calculation of the absorbed flux. The colored lines are the average model line profiles at the times of the average measured absorption (see \autoref{sec:sec6}). The fourth frame in each row shows the in--transit empirical Monte Carlo (EMC) $W_\lambda$ distributions for each line. Each absorption measurement is detected above the 3$\sigma$ level. The fifth column shows the EMC for the average pre--transit signal.}
\end{figure*}

Here, we present a time--resolved pre--transit signal detected in the high--resolution transmission spectra of the Balmer lines 
H$\alpha$, H$\beta$, and H$\gamma$ of HD 189733b. The observations and data reduction procedures are given in \autoref{sec:sec2}.
The transmission spectrum analysis and our procedure for estimating uncertainties in the absorption signal is presented in \autoref{sec:sec3}. 
In \autoref{sec:sec4} and \autoref{sec:sec5} we discuss the absorption signals and the effect of star spots and active regions.
The bow shock and exosphere models are described in \autoref{sec:sec6}. We discuss the planetary magnetic field strength estimate in \autoref{sec:sec7} and 
provide a brief conclusion in \autoref{sec:sec8}.

\section{OBSERVATIONS AND DATA REDUCTION}
\label{sec:sec2}

Two transits of HD 189733b were obtained on the two half nights of June 3 and July 4, 2013 using HiRES on Keck I \citep{vogt}. The approximate resolving power of the observations is $R$$\sim$68,000, or 4.4 km s$^{-1}$ at H$\alpha$. The B2 decker was employed which has a slit size of 7.0$^{\prime\prime}\times0.57^{\prime\prime}$. Exposure times ranged from 3 to 5 minutes. The average signal-to-noise of the extracted spectra is 400 at H$\alpha$, 180 at H$\beta$, and 120 at H$\gamma$.

The observations from June 3 begin at first contact and so add no significance to the pre-transit detection. In addition, the cadence of the June observations was twice as long, due to intermittent terrestrial clouds and haze, providing less detail on the shape of the transit. For these reasons we do not include the June data in the analysis. We note that the in-transit absorption values measured for the June data are similar to the values measured from the July data. 

The data were reduced using the HiRES Redux package written by Jason X. Prochaska\footnote{http://www.ucolick.org/$\sim$xavier/HIRedux/}. Standard reduction steps were taken including bias subtraction, flat fielding, and the removal of cosmic rays and hot pixels. The spectra were extracted using a 6.6$^{\prime\prime}$ boxcar. All images were examined manually for order overlap in the blue chip. HiRES Redux also performs 2D wavelength solutions using Th-Ar lamp exposures taken at the beginning of each night. The residuals for the wavelength fits are $\sim$0.05 pixels, or $\sim$0.02 \AA, in all orders. When applied to an individual observation, all wavelength solutions are corrected for Earth's heliocentric velocity and for the radial velocity of HD 189733. This places the spectrum in the rest frame of the star.

Telluric absorption is removed from the H$\alpha$ spectra using the telluric fitting program Molecfit \citep{kausch}. We first remove the strong H$\alpha$ line absorption from a telluric standard. The telluric absorption in the standard is then fit using Molecfit. The best telluric fit is then shifted appropriately in wavelength, scaled, and subtracted from each individual observation. This process results in transmission spectra with residuals in the telluric lines of $\sim$0.25\%. Telluric absorption from Mauna Kea is negligible in the H$\beta$ and H$\gamma$ orders. Telluric removal is not performed for these lines. 

\section{TRANSMISSION SPECTRUM AND EMC ANALYSIS}
\label{sec:sec3}

The transmission spectrum is defined as 

\begin{equation}
S_T=\frac{F_{i}}{F_{out}}-1
\end{equation}

\noindent where $F_{i}$ is a single observation and $F_{out}$ is the master post-transit spectrum. In order to obtain the final transmission 
spectra all observations must be normalized and aligned in wavelength space. Changes in spectral resolution due to temperature or instrument flexure are minimal and do not affect the spectra over the course of one half night. All spectra are initially normalized using a 4th order polynomial. A single out-of-transit spectrum is then chosen as the comparison spectrum in order to apply a wavelength correction to all spectra. A number of stellar absorption lines in each spectrum are cross correlated with the same lines in the comparison spectrum. We tested three different types of fits to the resulting wavelength shifts: constant, linear, and spline fits. The spline and linear fits perform no better, and in some cases worse, than a constant offset in wavelength. Thus the constant offsets are used to rectify the wavelength vectors for each spectrum relative to the comparison spectrum. This results in precisely aligned spectra where the largest residuals occur across 1--2 pixels at the cores of narrow spectral lines. Cores of identified stellar absorption lines in the normalized spectra are ignored when calculating the absorbed flux in each line so these residuals do not contribute to the transmission signal. Once spectra have been normalized and wavelength shifted they are divided by the master out-of-transit spectrum, a co-addition of the selected out-of-transit spectra. The resulting spectrum is renormalized to remove slowly varying continuum shapes leftover from the division. This occurs because the original normalization is slightly different for each spectrum. The line of interest is ignored when determining the renormalization so as to avoid introducing offsets from the true continuum. 

\autoref{fig:fig1} shows the average transmission spectrum for the pre-transit (first column), in-transit (second column), and post-transit
(third column) observations for H$\alpha$ (top row), H$\beta$ (middle row), and H$\gamma$ (bottom row). The model line profiles are
over plotted with solid lines (see \autoref{sec:sec4}). The number of spectra used 
to create each spectrum is give in the bottom-right of the top-row windows. The fourth and fifth columns shows the results of an empirical Monte Carlo
(EMC) process that we utilize to produce estimates for the absorption uncertainties \citep{redfield,jensen11,wyttenbach}. The EMC distributions are given in terms of 
$W_\lambda$, the equivalent width of absorption across the line. $W_\lambda$ is calculated by integrating the line profile from $-200$ to $+200$
km s$^{-1}$. Values of $W_\lambda$ do not change by more than 25\%, for any measurement, if the integration width is extended to $\pm$500 km s$^{-1}$. 

The in--transit EMC distributions are generated by randomly selecting 5000 subsets 
of in-transit spectra and comparing them to either the master out-of-transit spectrum (the In-Out method) or the master in-transit spectrum
(the In-In method). The standard deviations of the resulting distributions are taken to be the 1$\sigma$ uncertainties for the reported 
in-transit absorption. Each line is clearly detected at $>$3$\sigma$. We note that propagated Poisson errors result in very similar estimates
($\sim$20\% less than the EMC estimates) for the absorption uncertainties in each line. We choose to use the uncertainty derived from the 
EMC due to its ability to highlight systematic errors, which can dominate the measurement if not taken into account.

The small number of pre--transit observations ($N$=4) does not allow sufficient combinations of spectra to be generated for the EMC process
described above. Instead, we compare the average pre--transit spectrum to all possible combinations of the nine post--transit spectra ($N$=511).
The standard deviation of the resulting distribution, shown in the fifth column of \autoref{fig:fig1}, is taken as the 1--$\sigma$ uncertainty
for the measured absorption. This process is also used to estimate an average uncertainty for the individual points in \autoref{fig:fig3}.    

\begin{figure*}[ht]\label{fig:fig2}
   \centering
   \includegraphics[scale=.65,clip,trim=30mm 20mm 0mm 30mm,angle=0]{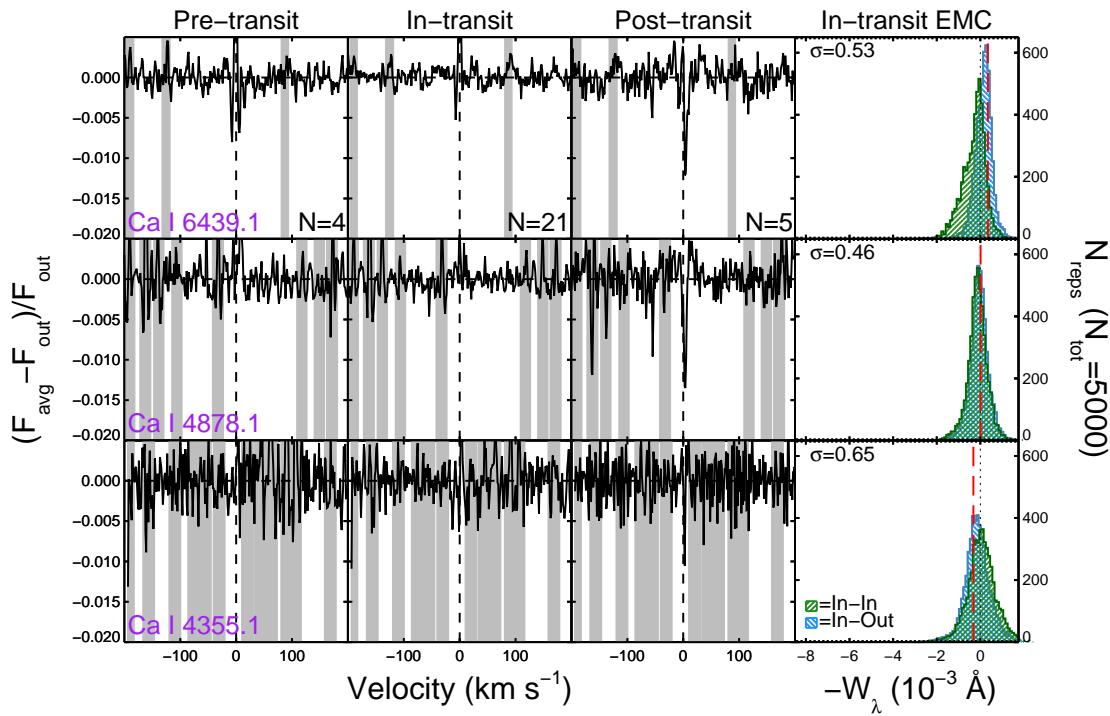} 
   \caption{Master line profiles for the Ca I control lines. The format is the same as \autoref{fig:fig1}. Each line is consistent 
   with no absorption indicating that systematics do not contribute to the absorption measured in the Balmer lines.}
\end{figure*}

As an additional test of the EMC procedure and to verify that we are actually measuring absorption in the Balmer lines, we perform the same in--transit EMC analysis for three control lines of Ca I. One line is chosen each for H$\alpha$, H$\beta$, and H$\gamma$ from the same spectral order that the Balmer line is located in. Ca I is expected to condense out of the atmospheres of hot Jupiter exoplanets and thus should not exist in atomic form in large quantities \citep{lodders}. If systematics from the reduction procedure are resulting in absorption in the Balmer lines we are likely to see similar artifacts in the Ca I control lines. The results of the control line transmission and EMC analyses are shown in \autoref{fig:fig2}. As expected, these lines show no evidence of absorption and the resulting uncertainties are similar to the uncertainties measured for the Balmer lines.   

\section{IN--TRANSIT ABSORPTION}
\label{sec:sec4}

The second column of \autoref{fig:fig1} shows the average in--transit absorption profiles for the July transit. The in--transit EMC is
shown in the fourth column. The absorbed fluxes measured in the line profiles of H$\alpha$, H$\beta$, and H$\gamma$ are all
detected at $>$3$\sigma$. The H$\beta$ and H$\gamma$ detections are the first reported measurements for an exoplanet 
atmosphere. The absorption that we measure is $\sim$50\% lower than that from \citet{jensen12}. This may be due to
variability in the surface features of HD189733 (e.g., spots and active regions). We model the in--transit profiles as arising
in the planetary exosphere (see \autoref{sec:sec62}). The possible contribution of star spots and active regions to the absorption
profile is discussed in \autoref{sec:sec51}.

\section{PRE--TRANSIT ABSORPTION}
\label{sec:sec5}

Absorption values, which are the negative of $W_\lambda$, determined from individual exposures are shown as a function of time in 
\autoref{fig:fig3}. The gap in the data from $\sim-115$ minutes until $-70$ minutes was, unfortunately, used to obtain telluric standards. The
strong pre-transit signal was not anticipated. The absorption is calculated by integrating the flux from $-200$ to $+200$ km s$^{-1}$. The error 
bars represent the average standard deviation of the EMC process using all combinations of the post--transit spectra. The average pre--transit
signal shown in \autoref{fig:fig1} is detected at $>3\sigma$ for H$\alpha$ and H$\beta$. The H$\gamma$ signal is detected slightly below the
3$\sigma$ level.    

The pre-transit absorption exhibits a few important properties. First, the ratio of the absorbed flux between H$\alpha$ and H$\beta$ is $\sim1.6-2.0$, which is much lower than in the optically thin limit. This indicates that the absorbing material is optically thick. Indeed, the line morphologies of the pre-transit observations appear saturated, i.e., flat bottomed and broad. Secondly, the absorption abruptly begins to decrease between $t-t_{mid}=-70$ and $-66$ minutes. This requires that the material causing the pre-transit absorption must begin to exit the stellar disk before the time of first optical contact (labeled $t_I$ in \autoref{fig:fig3}). This suggests that the distribution of absorbing material is asymmetric with respect to the planet, a point that is supported by the lack of any significant absorption seen at similar times post-transit \citep[in contrast to GJ 436, for example;][]{kulow}. Finally, the integrated absorption never exceeds $\sim$1.3\% which requires that the optically thick absorbing material cover a similar fraction of the stellar surface.

\subsection{The Effects of Star Spots, Active Regions, and Stellar Activity}
\label{sec:sec51}

Star spot crossings are not unusual for HD 189733b \citep{pont}. The transit of a star spot or active region can 
change the relative amount of stellar light being absorbed by the occulting material. The transit of a spot-free
chord can also change the relative amount of stellar light being absorbed, even for specific spectral features, if 
there is a non-negligible fraction of the stellar surface covered by spots \citep{berta}. Stellar activity can also mimic 
transit features when comparing spectra from different times: if the star brightens or dims relative to the time of
comparison, this signal can appear as the absorption of stellar flux when in truth it is merely variation in the amount
emitted by the star. For our case, stellar variability due to time-dependent Balmer line emission from active regions
can mimic the spectral line absorption: if the star is more active during the times when the out-of-transit spectra 
are measured, the filling in of the stellar Balmer line cores will appear as an absorption signature for any spectrum
measured at a time when the star is less bright. 

There is ample evidence that the absorption signatures we measure are not caused by
stellar variability. Both pre- and in-transit spectra are compared to
the same out-of-transit spectra. This requires that any absorption signal caused purely by stellar variability
have the same absorption strength in both the pre- and in-transit spectra. This is clearly not the case. In fact,
the pre-transit measurements are significantly stronger than the in-transit measurements despite the fact that
the exosphere, by way of the occulting planet, transits an effectively smaller stellar disk than the bow
shock. If the pre-transit absorption is entirely due to stellar variability, the in-transit absorption would necessarily be
\textit{stronger} than the pre-transit absorption. Since this relationship is not seen in the data, the absorption
cannot be solely due to stellar Balmer line variability. This means that the pre-transit signal must be
caused by absorbing material occulting the stellar disk. 

An indicator of stellar activity level is the $S$-index
measured from the Ca II H and K lines \citep{duncan}. The index, shown in \autoref{fig:fig4} for the July transit, has similar values
pre-transit compared to post-transit, although there appears to be a decrease about halfway through
the transit. The similar $S$-index values both before and after transit also indicate that the pre-transit absorption
is not caused by variable stellar emission.

The strength of spectral features can be affected by the relative contributions of the stellar photosphere
and spots or active regions \citep{berta}. While the transiting bow shock, which is transparent to continuum photons,
 is not subject to this affect, the opaque transiting planet certainly is. If the planet transits a spot-free chord, the 
 stellar spectrum will be weighted towards the spot spectrum and the relative strength of spectral features between 
 in and out-of-transit observations can change, even at the $1-2\%$ level. As a result, this effect is capable of
 producing spurious in-transit features. We do not believe this to be the case for our in-transit absorption
 measurements. In general, spotted active regions will fill in the Balmer line cores with emission, causing
 the line strength to effectively decrease. When the planet is transiting, this produces weaker absorption lines
 since the observed spectrum is weighted towards the active spotted regions. When compared to out-of-transit
 measurements, this will produce in-transit \textit{emission} features. The fact that we see the opposite, i.e.,
 strong in-transit absorption features, indicates that we are measuring absorption through the atmosphere 
 of the planet. However, the absolute strength of the measured absorption is almost certainly affected. The
 contribution of spotted active regions most likely causes us to underestimate the strength of the in-transit
 exospheric absorption. This may be the cause of the difference between the in-transit absorption strength
 measured here and that of \citep{jensen12}, who measure the in-transit line depth to be two times deeper.     
    
The arguments given above apply to the general features of the absorption time series. It is difficult, however,
to determine how individual absorption measurements are affected by a nonuniform stellar disk. Some absorption
measurements may well be affected by spots or bright regions on the stellar surface. For example, in between 
3rd and 4th contact the absorption varies between the expected values from the exiting exosphere and much
stronger absorption not accounted for by the model. These points could be due to the transitory occultation by
the exosphere of bright chromospheric regions on the limb of the star. Modeling of these variable features is beyond
the scope of this study but they provide a plausible explanation for the departure of individual points from the 
model estimates.

\section{Modeling the Absorption}
\label{sec:sec6}

We model the pre--transit absorption as arising in a thin bow shock orbiting ahead HD 189733 b. The in--transit absorption 
is modeled as arising in the exosphere of the planet. We neglect limb darkening in our model calculations since we are measuring the absorption across a very narrow ($\sim$1.0 \AA) portion of the stellar line. This portion of the Balmer lines forms in a very thin layer in the stellar atmosphere with a small temperature gradient and thus suffers negligibly from limb darkening \citep{mihalas}. 

\subsection{Bow shock model}
\label{sec:sec61}

We construct a simple model of the pre-transit absorption as arising in a bow shock orbiting ahead of the planet. While a bow shock is not the only possible cause of the absorption, it is a plausible one and has been successful at explaining the pre-transit UV observations of WASP12-b and HD 189733b \citep{llama11,benjaffel}. We do not attempt to model the full physical conditions of the stellar wind/corona and the resulting bow shock conditions. Instead, we assume a bow shock geometry and require that the $n=2$ hydrogen density assigned to the bow shock produce the necessary absorption. We assume that the standoff distance of the magnetosphere, $r_m$, is equal to the distance of the nose of the shock from the planet. 

The bow shock model is constructed using the shock geometry of \citet{wilkin}. A density is assigned to the nose of the shock and then scaled along the bow according to the scaling law 

\begin{equation}
\rho\left(r\right)=\rho_0\left(\frac{r_m}{r}\right)^\alpha
\end{equation}

\noindent where $r$ is the linear distance from the planet to the bow shock and $\alpha$ is the exponent describing how quickly the density changes with distance from the density at the nose, $\rho_0$. The distance $r$ is always greater than $r_m$ so the density decreases monotonically from the nose. A 3--D density grid is then filled according to the prescription of the density law with the chosen value of $\alpha$. The bow shock is rotated around the z--axis with angle $\theta_{sh}$ from the tangent to the planet's orbital motion. The column density of the material covering the stellar disk is then calculated. The cell size in the grid is 0.01 $R_p$, the same size as the shock thickness. We find that a cell size of this magnitude is required in order to reproduce the relatively small absorption values from the data. Larger cells, which necessarily result in a larger shock width, result in too much of the stellar disk being covered by optically thick material. We find values of $W_\lambda$ to within $\sim$5\% of the reported values if we decrease the grid cell size to 0.005 $R_p$, indicating that the shock thickness of 0.01 $R_p$ is not an artifact of the resolution. 

\begin{figure*}[htbp]\label{fig:fig3}
   \centering
   \includegraphics[scale=.65,clip,trim=25mm 20mm 25mm 20mm,angle=0]{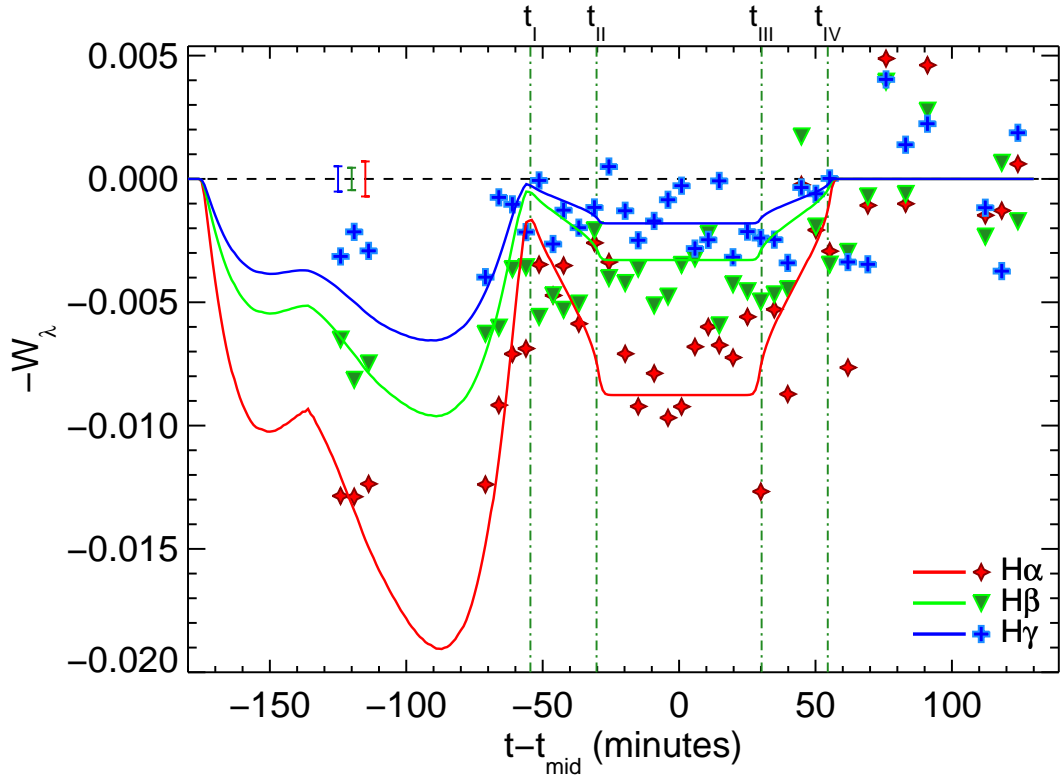} 
   \caption{Absorption as a function of time from transit midpoint for a single transit of HD 189733b. First through fourth contact are marked with the vertical green dashed-dotted lines. A sharp decrease in the absorbed flux can be seen from $-70$ min$<t<-40$ min. The gap in the 
   data from $-115$ minutes to $-70$ minutes was used for telluric standard observations. The absorption does not appear post-transit. The model is shown with solid lines. The uncertainties in the absorbed flux for each individual spectrum are shown in the upper-left. Each bar shows the average of the standard deviations of $W_\lambda$ obtained from the EMC for each point.}
\end{figure*}

 \begin{figure}[htbp]\label{fig:fig4}
   \centering
   \includegraphics[scale=.40,clip,trim=30mm 20mm 0mm 20mm,angle=0]{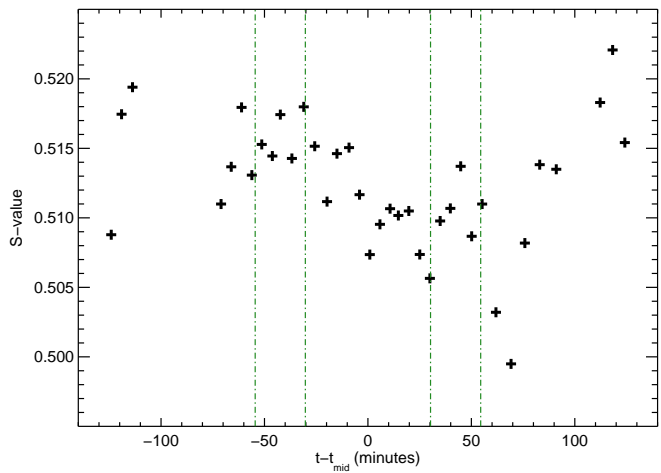} 
   \caption{Ca II H and K $S$--index measurements taken simultaneously with the Balmer lines. Propagated Poisson
   uncertainties are smaller than the plot symbols. Transit contact times are marked with the vertical green lines.
   There is no direct correlation with the transit, although there appears to be a decrease in the activity index at $t-t_{mid}=0$ minutes.}
\end{figure}

The optical depth in the Balmer lines is approximated by a Doppler-broadened delta function, $\tau_v=\tau_0 e^{-(v/b)^2}$, where $b=\sqrt{2}\sigma_v$ and $\sigma_v$ is the dispersion of a 1--D Gaussian velocity distribution \citep{draine}. The optical depth at line center is 

\begin{equation}
\tau_0=\frac{\sqrt{\pi}e^2f_{lu}N_l\lambda_{lu}}{m_e c b} 
\end{equation}

\noindent where $f_{lu}$ is the oscillator strength of the transition, $\lambda_{lu}$ is the central wavelength, $m_e$ is the electron mass, $e$ is the electron charge, $c$ is the speed of light in vacuum, and $N_l$ is the column density of the lower energy level of the transition. We use the optical depth as a function of velocity from line center to approximate an absorption line profile for each grid point using the column density calculated from the 3D bow shock. The final line profile for a single model iteration (i.e., a single time of the bow shock transit) is the sum of each individual line profile for each grid point on the stellar disk. Examples of the pre-transit model line profiles are overplotted on the measured spectra in \autoref{fig:fig1} We find that $b=4.1$ km s$^{-1}$ is able to adequately reproduce the line profile shapes. The final model $W_\lambda$ values are calculated identically to the measured values of $W_\lambda$. We point out that the model seems to poorly reproduce the pre-transit H$\gamma$ line profile and absorption values. We believe this is due to the blending of H$\gamma$ with a nearby Fe I line which causes the red wavelength side of the H$\gamma$ profile to show weaker absorption. This can be seen in \autoref{fig:fig1} where the model line profile matches the blue side of the line very well. A similar effect is seen for the in-transit H$\gamma$ profiles. Based on the line profile shapes of H$\alpha$ and H$\beta$ we have no reason to believe the shallow red-ward portion of the H$\gamma$ line is representative of anything physical. 

\begin{figure}[ht]\label{fig:fig5}
   \centering
   \includegraphics[scale=.48,clip,trim=15mm 15mm 10mm 50mm,angle=0]{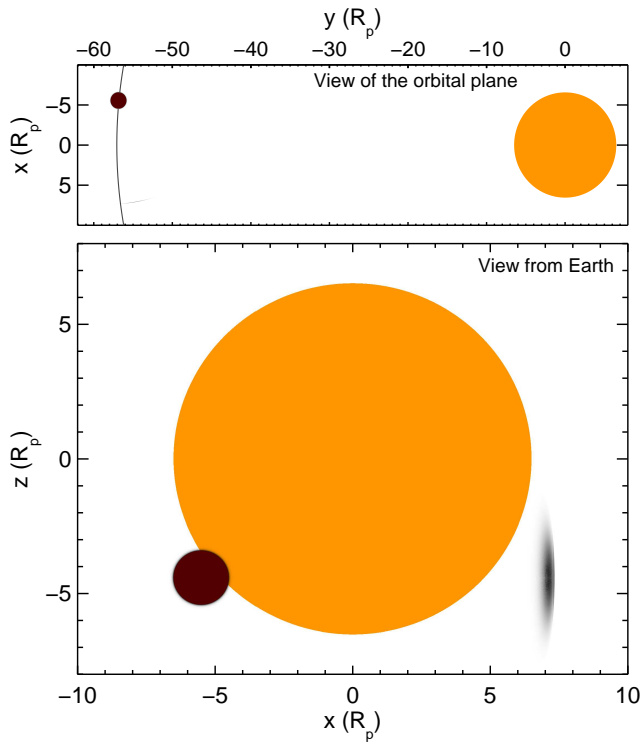} 
   \caption{To-scale projections of the planet and bow shock in the orbital plane (top panel) and the view from Earth (bottom panel) for time $t-t_{mid}=-50$ minutes, soon after the bow shock has exited the stellar disk. The planet is in between first and second contact. The exosphere and bow shock are scaled to the same density. The density in the bow shock can be seen to fall off rapidly away from the nose of the shock.}
\end{figure}

The model absorption values are shown as solid lines in \autoref{fig:fig3} and the model line profiles are shown as solid lines in \autoref{fig:fig1}. A to-scale snapshot of the system is shown in \autoref{fig:fig5}. The large standoff distance and the observed absorption time series are representative of the double transit suggested by \citet{vidotto11}. This model was calculated using values of $r_m$=12.75 $R_p$, $\theta_{sh}=15^\circ$, the angle between the planet's trajectory and the nose of the shock, and $\rho_0=9\times$10$^{-20}$ g cm$^{-3}$, the mass density of excited hydrogen at the nose of the shock. A scaling factor of $\alpha=400$ is required to sufficiently concentrate the density at the nose which is necessitated by the loss of absorption near t$_I$. The model reproduces the main features of the observed pre-transit time series: 1. the approximate ratio of H$\alpha$ to H$\beta$ absorption; 2. similar absorption values at $-125$ and $-70$ minutes; and 3. the steep decrease in the absorption immediately after $t-t_{mid}=-70$ minutes. The model also predicts an absorption maximum at $t-t_{mid}$$=-85$ minutes and ingress of the bow shock at $t-t_{mid}$$=-175$ minutes.

The model parameters reported here are fairly tightly constrained. For example, setting $\theta_{sh}=40^\circ$ results in too large of a projected area of the bow shock and requires the density to become even more highly concentrated at the nose. Similarly, smaller values of $\alpha$ do not concentrate the material enough at the nose and the absorption values are too large. The standoff distance must also increase to allow the shock to exit the stellar disk at $t-t_{mid}=-70$ minutes. We note that the values of $r_m=12.75$ $R_p$ and $\theta_{sh}=15^\circ$ derived in our favored model are similar to those found by \citet{benjaffel} (16.7 $R_p$ and 10-30$^\circ$, respectively) for their bow shock model of a pre-transit C II absorption signal. However, a value of $r_m=16.7$ $R_p$ is too large for our data since this would imply egress of the bow shock from the stellar disk earlier than observed, assuming a similar value of $\theta_{sh}$. Variability related to the stellar wind and/or corona, which has likely been observed to cause variations in the exospheric absorption signature from HD 189733b \citep{desetangs12}, could account for this difference. Thus we believe our values represent a plausible scenario for producing the measured absorption.

The incoming material in a bow shock is not stopped by the magnetosphere: momentum is conserved and the material flows around the bow \citep{wilkin}. 
Our model does not incorporate the bulk flow of material along the bow. The excited hydrogen density we find from our model is thus representative of
the physical state of the gas at the nose of the shock. The rapid falloff of the density from the nose implies that the shocked material cools as it moves
away from the nose and that the increase in temperature and density caused by the shock at distances far from the nose is insufficient to produce a significant 
amount of excited hydrogen.  

The large values of $r_m$ and small values of $\theta_{sh}$ found here and in the study of \citet{benjaffel} contrast with the relatively small values of $r_m$ ($\sim$3.8 $R_p$) and large values of $\theta_{sh}$ ($\sim$65--70$^\circ$) predicted by \citet{llama13} using a full 3D MHD simulation of the stellar wind from HD 189733 and assuming a dipolar magnetic field strength of 14 G at the pole. The large discrepancies suggest that these simulations overestimate the stellar wind speed and/or temperature or that the planetary magnetic field is underestimated.

The physical conditions in the bow shock are expected to fluctuate as the stellar wind speed, density, and temperature likely vary with time. For example, as the stellar wind speed increases the angle $\theta_{sh}$ will increase, moving the nose of the shock closer to the star-planet line. The low value of $\theta_{sh}=15^\circ$ reported here indicates that the planet is moving through a relatively slow moving wind, $\sim$30--40 km s$^{-1}$, or perhaps a hydrostatic corona if the true angle is even smaller \citep{vidotto10}. This may indicate that the planet is moving through a quiet or slow region in the stellar wind. Assuming that the planetary magnetic field strength remains constant, the standoff distance $r_m$ will also change depending on the varying thermal and ram pressure from the stellar wind. Even if the conditions for a shock to form are not met, the incoming stellar wind may still undergo some compression and heating as it is slowed by the magnetosphere \citep{lai}. Fluctuations in the pre-transit signal could be used as a probe of stellar wind variability. 

As a first attempt at using our measurement to estimate properties of HD 189733's wind, we can use our derived model parameters to calculate a mass loss rate for the star.
If we assume a spherical outflow of constant velocity, the density we estimate at the orbital distance of HD 189733b translates into a density at the stellar surface of
$\rho_0$=1.8$\times$10$^{-18}$ g cm$^{-3}$. Assuming a wind velocity of 40 km s$^{-1}$, we estimate a mass loss rate of $\dot{M}$=2.5$\times$10$^{11}$ g s$^{-1}$,
or $\sim$4$\times$10$^{-15}$ $\Msun$ yr$^{-1}$. This value is $\sim$100$\times$ lower than that predicted by \citet{llama13} (4.5$\times$10$^{-13}$ $\Msun$ yr$^{-1}$). 
This suggests that a small fraction of the stellar wind material ends up as excited hydrogen in the bow shock.

This estimate is a lower limit: the density from our model is the density of the excited hydrogen in the bow shock
and does not include ground state hydrogen or other elements. This mass loss estimate has other obvious limitations. For example, the stellar wind is likely
not spherically symmetric. In addition, this estimate is essentially a snapshot in time. HD 189733 is an active star and likely has a variable wind. Our estimate 
is also lower than the mass loss rate suggested by the X--ray flux versus $\dot{M}$ relationship given in \citet{wood14}. Using an average X--ray flux of 
$F_X$=6.3$\times$10$^5$ taken from the NEXXUS2 catalog\footnote{http://www.hs.uni-hamburg.de/DE/For/Gal/Xgroup/nexxus/},
we calculate a mass loss rate of $\dot{M}$=1.1$\times$10$^{-12}$ $\Msun$ yr$^{-1}$. Although the \citet{wood14} relationship is fairly uncertain, this value
is more in line with the \citet{llama13} prediction than the value estimated from our model. Thus the low mass loss rate estimated here should be
considered a rough lower limit for HD 189733. Future observations will help constrain the variability and potentially measure changes in the stellar 
mass loss rate.

\subsection{Exosphere model}
\label{sec:sec62}

The exosphere model consists of an isothermal gas in hydrostatic equilibrium above the planet. We calculate the density in the exosphere out to 10$H_{exo}$ where $H_{exo}$ is the exospheric scale height given by $H_{exo}=k_b T/\left(m_h g\right)$ where $k_b$ is Boltzmann's constant, $T$ is the exospheric temperature, and $g$ is the acceleration due to gravity at the planetary surface. The exobase is taken to be at $R_p$=1.0. In order to roughly match the in-transit absorption, we find $H_{exo}=0.04$ $R_p$ and $T_{exo}=7,000$ K with a base number density of $n_0=300$ cm$^{-3}$. This simple model produces systematically low estimates for the H$\beta$ absorption. This likely indicates that the exosphere does not follow a simple hydrostatic pressure law and a more realistic treatment is necessary to more closely match the line ratios. We note that the exospheric parameters derived here and very similar to those derived by \citet{christie} who give a detailed treatment of the $n=2$ hydrogen level population in the exosphere of HD 189733b. 

\section{ESTIMATING THE PLANETARY MAGNETIC FIELD STRENGTH}
\label{sec:sec7}

Our measurements can be used to estimate the planetary magnetic field strength by considering some limiting cases of pressure balance between the interplanetary material (i.e., stellar wind or corona) and the planetary magnetosphere. Rewriting equation (9) from \citet{llama13} we see that the magnetic field strength of the planet at the pole is

\begin{equation}
B_p = 2 \left( \frac{r_m}{R_p} \right)^3 \left[8\pi\left(\rho_w \Delta u_w^2+P_w\right)+B_w^2\right]^{1/2}
\end{equation} 

\noindent where $\rho_w$ is the mass density of the stellar wind or corona, $\Delta u_w$ is the relative velocity of the planet and the wind, $P_w$ is the thermal pressure of the wind, and $B_w$ is the interplanetary magnetic field strength (i.e., the stellar magnetic field strength at the orbital radius of the planet). All of the pressure terms are the values at the orbital radius of the planet. Equation 4 neglects any pressure from the planetary atmosphere or wind. The equatorial magnetic field strength, assuming the field is purely dipolar, is one-half the polar value.

We adopt our favored model parameters and assume that the bow shock is mediated by a magnetosphere. If we neglect the pressure due to the stellar magnetic field, the dipolar magnetic field strength of the planet can be estimated by assuming pressure balance between the stellar wind and the planetary magnetosphere \citep{vidotto10,llama13}. By neglecting the stellar magnetic pressure we are essentially estimating a lower limit for the field since any addition to the pressure will require a larger planetary field strength for the inferred standoff distance. We choose a stellar wind temperature of $T_w=1\times10^{6}$ K which is roughly consistent with a wind of velocity of $<$100 km s$^{-1}$ at 0.03 AU \citep{matsakos}. The wind is assumed to be traveling radially at 40 km s$^{-1}$. The azimuthal wind velocity is neglected. The stellar wind is taken to be an ideal gas composed purely of protons to provide an estimate for $P_w$. We also assume that the material in the shock has been compressed by a factor of 4, i.e., the shock is adiabatic, resulting in a stellar wind density of $\rho_w$=2.5$\times$10$^{-20}$ g cm$^{-3}$. We note that this density is very similar to the stellar wind density derived by \citet{bourrier13a} ($\sim$3$\times$10$^{-20}$ g cm$^{-3}$) to reproduce observed L$\alpha$ absorption in HD 189733b.

Given these assumptions we estimate an equatorial magnetic field strength of $B_{eq}=28$ G for HD 189733b, similar to the upper limit of 24 G derived by \citet{vidotto10} for WASP12-b. The estimated field strength is $\sim$7 times larger than Jupiter's equatorial field ($\sim$4 G) and is driven by the large value of $r_m$. The scaling relation of \citet{reiners09} is used by \citet{reiners10} to estimate the magnetic field strength of HD 189733b at $\sim$14 G. The factor of two difference between our estimates and those from the scaling relation indicate a factor of $\sim$8 difference in the internal heat flux used in the scaling relation estimate. The additional heat source suggested by our estimates could have significant implications for models of hot Jupiter interiors. 

While we have focused on a planetary bow shock as the mechanism for producing the pre-transit absorption, alternate scenarios exist for producing similar signals. The effects of charge exchange between a hot stellar wind and a planetary wind are investigated in \citet{tremblin}. Their model, which has a similar geometry to the bow shock approximation, produces a layer of hot, dense neutral hydrogen at a similar standoff distance ($\sim$10 $R_p$) as found in our model and shows decent agreement with Lyman $\alpha$ absorption measurements. Their model likely predicts too large of an excited hydrogen density over too large a volume to match the low absorption values found here. However, a more detailed investigation of excited hydrogen absorption in this context would be informative. Indeed, charge exchange may be occurring in the bow shock between stellar wind protons and neutral hydrogen that has escaped from the planet, contributing to the $n$=2 hydrogen density. Transiting accretion streams from the planet, which is overflowing its Roche lobe, to the star can produce sufficient column density and coverage of the stellar disk to result in pre-transit absorption \citep{lai}. This scenario, however, should produce red-shifted absorption which is not observed in our data. In fact, we find no absorption centroids shifted by more than $\pm$5 km s$^{-1}$ from line center. This suggests that the excited hydrogen has a symmetric bulk velocity along the line of sight.

\section{SUMMARY AND CONCLUSIONS}
\label{sec:sec8}

We have presented our detection of resolved pre-transit line absorption around HD 189733b in the first three members of the hydrogen 
Balmer series. A bow shock model is able to reproduce the important features of the absorption time series. The physical parameters 
derived from our model suggest an equatorial planetary magnetic field strength of $B_{eq}=28$ G. This large value is driven mainly by the 
large standoff distance of the magnetosphere $r_m=12.75$ $R_p$, a parameter that is well constrained by the model. For the
derived standoff distance, this field strength is likely a lower limit due to the multiple simplifying assumptions employed in the estimate.
More detailed models which include the stellar magnetic field pressure and a full treatment of the stellar wind will prove insightful. 
The expected variability of the bow shock signal can also be tested by observing multiple transits across many epochs. Observations
of this type could be useful probes of stellar wind variability for HJ hosts.

The magnetic field of a HJ plays an important role in determining the planetary mass loss rate: the stronger the magnetic field, the lower the outflow rate \citep{owen}. Planetary magnetic fields also provide a potential source of energy for SPIs via reconnection events with the stellar corona \citep{cuntz}. This energy flux scales as $B_{pl}$ \citep{cuntz,lanza} so larger magnetic field strengths can potentially induce more energetic SPIs. If the field strengths estimated above are common for HJ exoplanets, this regime needs to be more fully explored by mass loss and SPI models which have generally considered field strengths similar to or less than that of Jupiter \citep{lanza,trammell,strugarek}.

Planetary magnetospheres are a revealing diagnostic of planetary interiors and the first line of defense against energetic stellar wind particles which can have a devastating impact on planetary atmospheres and life near the surface of the planet.  While considered to be a critical property in understanding a planet, its evolution, and habitability, the prospects of measuring an exoplanetary magnetic field were thought to be remote. However, if independent measurements can be made of the planet hosts' stellar wind, observations such as those presented here may provide important constraints on exoplanetary magnetic fields. The large field strength we estimate implies that lower magnetic field strengths, as found in our solar system, may not be ubiquitous among other exoplanet systems.  A population study of magnetic field properties of exoplanets that display bow shock-like structures will complement measurements of solar system magnetospheres and be used to better understand planetary interiors and the intertwined relationship of magnetic fields, planetary atmospheres, and habitability.

\bigskip

{\bf Acknowledgments:} The data presented herein were obtained at the W.M. Keck Observatory, which is operated as a scientific partnership among the California Institute of Technology, the University of California and the National Aeronautics and Space Administration. The Observatory was made possible by the generous financial support of the W.M. Keck Foundation. The authors wish to recognize and acknowledge the very significant cultural role and reverence that the summit of Mauna Kea has always had within the indigenous Hawaiian community.  We are most fortunate to have the opportunity to conduct observations from this mountain. Keck telescope time was granted by NOAO, through the Telescope System Instrumentation Program (TSIP) (Program ID 2013A-0174, PI: A. Jensen). TSIP is funded by NSF. This work was completed with support from the National Science Foundation through Astronomy and Astrophysics Research Grant AST-1313268 (PI: S.R.). The authors wish to express gratitude to the observers J. Brewer and J. Moriarty for their crucial help obtaining the data. The authors also extend their thanks to Wesleyan student observers E. Edelman, C. Malamut, R. Martinez, and B. Tweed for their participation. P. W. C. and S. R. acknowledge useful and stimulating conversations with M. Swain.


\begin{thebibliography}{}

\bibitem[Ben-Jaffel \& Ballester(2013)]{benjaffel}Ben-Jaffel, L., \& Ballester, G. E. 2013, A\&A, 553, A52

\bibitem[Berta et al.(2011)]{berta}Berta, Z. K., Charbonneau, D., Bean, J., et al. 2011, ApJ, 736, 12

\bibitem[Bouchy et al.(2005)]{bouchy}Bouchy, F., Udry, S., Mayor, M., et al. 2005, A\&A, 444, L15

\bibitem[Bourrier et al.(2013)]{bourrier13}Bourrier, V., Lecavelier des Etangs, A., Dupuy, H., et al. 2013, A\&A, 551, A63

\bibitem[Bourrier \& Lecavelier des Etangs(2013)]{bourrier13a}Bourrier, V., \& Lecavelier des Etangs, A. 2013, A\&A, 557, A124

\bibitem[Burrows et al.(2007)]{burrows}Burrows, A., Hubeny, I., Budaj, J., \& Hubbard, W. B. 2007, ApJ, 661, 502

\bibitem[Christie et al.(2013)]{christie}Christie, D., Arras, P., \& Li, Z.-Y. 2013, ApJ, 772, 144

\bibitem[Cuntz et al.(2000)]{cuntz}Cuntz, M., Saar, S. H., \& Musielak, Z. E. 2000, ApJ, 533, L151

\bibitem[Draine(2011)]{draine}Draine, B. T. 2011, \textit{Physics of the Interstellar and Intergalactic Medium} (Princeton University Press, Princeton, NJ)

\bibitem[Duncan et al.(1991)]{duncan}Duncan, D. K., Vaughan, A. H., Wilson, O. C., et al. 1991, ApJS, 76, 383

\bibitem[Fossati et al.(2010)]{fossati}Fossati, L., Haswell, C. A., Froning, C. S., et al. 2010, ApJ, 714, L222

\bibitem[Grillmair et al.(2008)]{grillmair}Grillmair, C. J., Burrows, A., Charbonneau, D., et al. 2008, Nature, 456, 767

\bibitem[Jensen et al.(2011)]{jensen11}Jensen, A. G., Redfield, S., Endl, M., et al. 2011, ApJ, 743, 203

\bibitem[Jensen et al.(2012)]{jensen12}Jensen, A. G., Redfield, S., \& Endl, M., et al. 2012, ApJ, 751, 86

\bibitem[Kausch et al.(2014)]{kausch}Kausch, W., Noll, S., Smette, A., et al. 2014, ASPC, 485, 403

\bibitem[Kulow et al.(2014)]{kulow}Kulow, J. R., France, K., Linsky, J., \& Loyd, R. O. P. 2014, ApJ, 786, 132

\bibitem[Lai et al.(2010)]{lai}Lai, D., Helling, Ch., \& van den Heuvel, E. P. J. 2010, ApJ, 721, 923

\bibitem[Lanza(2013)]{lanza}Lanza, A. F. 2013, A\&A, 557, A31

\bibitem[Lecavelier des Etangs et al.(2010)]{desetangs}Lecavelier des Etangs, A., Ehrenreich, D., Vidal-Madjar, A., et al. 2010, A\&A, 514, A72

\bibitem[Lecavelier des Etangs et al.(2012)]{desetangs12}Lecavelier des Etangs, A., Bourrier, A., Wheatley, P. J., et al. 2012, A\&A, 543, L4

\bibitem[Llama et al.(2011)]{llama11}Llama, J., Wood, K., Jardine, M. et al., 2011, MNRAS, 416, L41

\bibitem[Llama et al.(2013)]{llama13}Llama, J., Vidotto, A. A., Jardine, M., et al. 2013, MNRAS, 436, 2179

\bibitem[Lodders(1999)]{lodders}Lodders, K. 1999, ApJ, 519, 793

\bibitem[Matsakos et al.(2015)]{matsakos}Matsakos, T., Uribe, A., \& K\"{o}nigl, A. 2015, A\&A, 578, A6

\bibitem[Mihalas(1978)]{mihalas}Mihalas, D. 1978, \textit{Stellar Atmospheres} (W. H. Freeman and Co., San Francisco) 

\bibitem[Owen \& Adams(2014)]{owen}Owen, J. E., \& Adams, F. C. 2014, MNRAS, 444, 3761

\bibitem[Pont et al.(2013)]{pont}Pont, F., Sing, D. K., Gibson, N. P., et al. 2013, MNRAS, 432, 2917

\bibitem[Redfield et al.(2008)]{redfield}Redfield, S., Endl, M., Cochran, W. D., \& Koesterke, L. 2008, ApJ, 673, L87

\bibitem[Reiners et al.(2009)]{reiners09}Reiners, A., Basri, G., \& Christensen, U. R. 2009, ApJ, 697, 373

\bibitem[Reiners \& Christensen(2010)]{reiners10}Reiners, A., \& Christensen, U. R. 2010, A\&A, 522, A13

\bibitem[Shkolnik et al.(2005)]{shkolnik}Shkolnik, E., Walker, G. A. H., Bohlender, D. A., Gu, P.-G., \& K\"{u}rster, M. 2005, ApJ, 622, 1075 


\bibitem[Strugarek et al.(2014)]{strugarek}Strugarek, A., Brun, A. S., Matt, S. P., \& R\'{e}ville, V. 2014, ApJ, 795, 86

\bibitem[Swain et al.(2008)]{swain08}Swain, M. R., Vasisht, G., \& Tinetti, G. 2008, Nature, 452, 329

\bibitem[Trammell et al.(2011)]{trammell}Trammell, G. B., Arras, P., \& Li, Z.-Y. 2011, ApJ, 728, 152

\bibitem[Tremblin \& Chiang(2013)]{tremblin}Tremblin, P., \& Chiang, E. 2013, MNRAS, 428, 2565

\bibitem[Vidal-Madjar et al.(2003)]{vidal}Vidal-Madjar, A., Lecavelier des Etangs, A., D\'{e}sert, J.-M., et al. 2003, Nature, 422, 143

\bibitem[Vidotto et al.(2010)]{vidotto10}Vidotto, A. A., Jardine, M., \& Helling, Ch. 2010, ApJ, 722L, 168

\bibitem[Vidotto et al.(2011)]{vidotto11}Vidotto, A. A., Jardine, M., \& Helling, Ch. 2011, MNRAS, 414, 1573 

\bibitem[Vogt et al.(1994)]{vogt}Vogt, S. S., Allen, S. L., Bigelow, B. C., et al. 1994, SPIE, 2198, 362 

\bibitem[Wilkin(1996)]{wilkin}Wilkin, F. P. 1996, ApJ, 459, L31

\bibitem[Wood et al.(2014)]{wood14}Wood, B. E., M\"{u}ller, H.-R., Redfield, S., \& Edelman, E. 2014, ApJL, 781, L33

\bibitem[Wyttenbach et al.(2015)]{wyttenbach}Wyttenbach, A., Ehrenreich, D., Lovis, C., Udry, S., \& Pepe, F. 2015, A\&A, 577, A62

\end{thebibliography}
\end{document}